\documentstyle[multicol,aps,epsf]{revtex}
\begin{document}
\title{Degree Distributions of Growing Networks}
\author{P.~L.~Krapivsky$^{1}$, G.~J.~Rodgers$^2$, and S.~Redner$^1$}
\address{$^1$Center for BioDynamics, Center for Polymer Studies, 
and Department of Physics, Boston University, Boston, MA 02215}
\address{$^2$Department of Mathematical Sciences, Brunel University, 
Uxbridge, Middlesex, UB8 3PH, UK}

\maketitle
\begin{abstract}
  The {\em in-degree} and {\em out-degree} distributions of a growing
  network model are determined.  The in-degree is the number of incoming
  links to a given node (and {\it vice versa} for out-degree).  The
  network is built by (i) creation of new nodes which each immediately
  attach to a pre-existing node, and (ii) creation of new links between
  pre-existing nodes.  This process naturally generates correlated in-
  and out-degree distributions.  When the node and link creation rates
  are linear functions of node degree, these distributions exhibit {\em
    distinct\/} power-law forms.  By tuning the parameters in these
  rates to reasonable values, exponents which agree with those of the
  web graph are obtained.

\end{abstract}
\begin{multicols}{2}
 
The world-wide web (WWW) is a rapidly evolving network which now
contains nearly $10^9$ nodes.  Much recent effort has been devoted to
characterizing the underlying directed graph formed by these nodes and
their connecting hyperlinks -- the so-called ``web''
graph\cite{huberman,klein,kum,brod}.  In parallel with these
developments, a variety of growing network models have recently been
introduced and
studied\cite{barabasi,jose,klr,doro,newman,bb,kr,stefan,simon,gen}.
These model networks are built by sequentially adding both nodes and
links in a manner which mimics the evolution of real network systems,
with the WWW being the most obvious example.
  
One fundamental characteristic of any graph is the number of links at a
node -- the node degree.  The growing network models cited above predict
that the distribution of node degree has a power law form for growth
rules in which the probability that a newly-created node attaches to a
pre-existing node increases linearly with the degree of the ``target''
node\cite{barabasi,jose,klr,doro}.  This power law behavior strongly
contrasts with the Poisson degree distribution of the classical random
graphs\cite{rg}, where links are randomly created between any pair of
pre-existing nodes in the network.
  
\begin{figure}
  \narrowtext \epsfxsize=2.2in \hskip 0.5in
\epsfbox{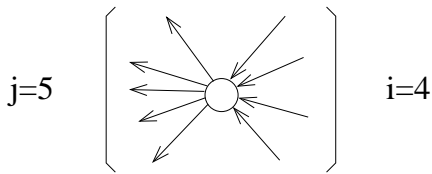} \vskip 0.1in
\caption{A node with in-degree $i=4$, out-degree $j=5$, and total
  degree 9.
\label{degrees}}
\end{figure}

Since web links are directed, the {\em total\/} degree of a node may
naturally be resolved into the {\em in-degree} -- the number of incoming
links to a node, and {\em out-degree} -- the number of outgoing links
from a node (Fig.~\ref{degrees}).  While the total node degree and its
distribution are now reasonably understood
\cite{barabasi,jose,klr,doro,kr}, little is known about the joint
distribution of in-degrees and out-degrees, as well as their
correlation.  Empirical measurements of the web indicate that in-degree
and out-degree distributions exhibit power-law behaviors with different
exponents\cite{klein,kum,brod}.  In this Letter, we solve for the joint
distribution in a simple growing network model.  We are able to
reproduce the observed in-degree and out-degree distributions of the web
as well as find correlations between in- and out-degrees of each node.

\begin{figure}
  \narrowtext \epsfxsize=2.9in \hskip 0.1in
\epsfbox{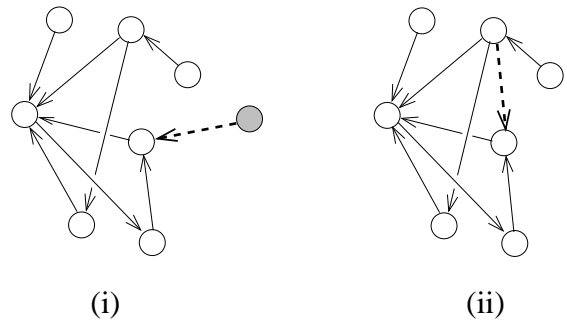} \vskip 0.1in
\caption{Illustration of the  growth processes in the growing network model: 
  (i) node creation and immediate attachment, and (ii) link creation.  In (i)
  the new node is shaded, while in both (i) and (ii) the new link is dashed.
\label{io-growth}}
\end{figure}

Our model represents an extension of growing network models with node and
link creation\cite{simon,gen} to incorporate link directionality.  The
network growth occurs by two distinct processes (Fig.~\ref{io-growth}):
\begin{itemize}
\item[(i)] With probability $p$, a new node is introduced and it immediately
  attaches to one of the earlier target nodes in the network.  The attachment
  probability depends only on the in-degree of the target.
\item[(ii)] With probability $q=1-p$, a new link is created between
  already existing nodes.  The choices of the originating and target
  nodes depend on the out-degree of the originating node and the
  in-degree of the target node.
\end{itemize}

If only process (i) was allowed, the out-degree of each node would be one by
construction.  Process (ii) has been shown to drive a transition in the
network structure\cite{gen}.  We shall further show that this general model
gives a non-trivial out-degree distribution which is distinct from the
in-degree distribution.

We begin our analysis by determining the average node degree; this can
be done without specification of the attachment and link creation
probabilities.  Let $N(t)$ be the total number of nodes in the network,
and let $I(t)$ and $J(t)$ be the total in-degree and out-degree,
respectively.  According to the two basic growth processes enumerated
above, at each time step these degrees evolve according to one of the
following two possibilities
\begin{equation}
\label{growth}
(N,I,J)\to \cases{(N+1,I+1,J+1)  & probability $p$,\cr
                  (N,I+1,J+1)    & probability $q$.}  
\end{equation}
That is, with probability $p$ a new node and new directed link are created
(Fig.~\ref{io-growth}) so that the number of nodes and both node degrees
increase by one.  Conversely, with probability $q$ a new directed link is
created and the node degrees each increase by one, while the total number of
nodes is unchanged.  As a result,
\begin{equation}
\label{NIJ}
N(t)=pt, \qquad I(t)=J(t)=t,
\end{equation}
from which we immediately conclude that the average in- and out-degrees,
${\cal D}_{\rm in}\equiv I(t)/N(t)$ and ${\cal D}_{\rm out}\equiv J(t)/N(t)$,
are both time independent and equal to $1/p$.

To determine the joint degree distributions, we need to specify: (i) the
{\em attachment rate} $A(i,j)$, defined as the probability that a
newly-introduced node links to an existing node with $i$ incoming and
$j$ outgoing links, and (ii) the {\em creation rate}
$C(i_1,j_1|i_2,j_2)$, defined as the probability of adding a new link
from a $(i_1,j_1)$ node to a $(i_2,j_2)$ node.  We restrict the form of
these rates to those which we naturally expect to occur in systems such
as the web graph.  First, we assume that the attachment rate depends
only on the in-degree of the target node, $A(i,j)=A_i$.  We also assume
that the link creation rate depends only on the out-degree of the node
from which it emanates and the in-degree of the target node, that is,
$C(i_1,j_1|i_2,j_2)= C(j_1,i_2)$.

On general grounds, the attachment and creation rates $A_i$ and $C(j,i)$
should be non-decreasing functions of $i$ and $j$.  For example, a web-page
designer is more likely to construct hyperlinks to well-known pages rather
than to obscure pages.  Similarly, a web page with many outgoing hyperlinks
is more likely to create even more hyperlinks.  We have found that the degree
distributions exhibit qualitatively different behaviors depending on whether
the asymptotic dependence of the rates $A_i$ and $C(j,i)$ on both $i$ and $j$
grow slower than linearly, linearly, or faster than linearly.  The first and
last cases lead to either rapidly decaying degree distributions or to the
dominance of a single node; this same behavior was already found for the
total node degree\cite{klr,kr}.  The most interesting behavior arises for
asymptotically linear rates, and we focus on this class of models in our
investigations.

Specifically, we consider the model with attachment and creation rates which
are shifted linear functions in all indices ({\em linear-bilinear} rates)
\begin{equation}
\label{rates}
A_i=i+\lambda, \qquad C(j,i)=(i+\lambda)(j+\mu).
\end{equation}
An intuitively natural feature of this model is that both the attachment
and creation rates have the same dependence on the popularity of the
target node.  The parameters $\lambda$ and $\mu$ in the rates of
Eq.~(\ref{rates}) must obey the constraints $\lambda>0$ and $\mu>-1$ to
ensure that the corresponding rates are positive for all permissible
values of in- and out-degrees, $i\geq 0$ and $j\geq 1$.

As the network grows, the joint degree distribution, $N_{ij}(t)$,
defined as the average number of nodes with $i$ incoming and $j$
outgoing links, builds up.  To solve for $N_{ij}(t)$, we shall use the
rate equation approach, which has recently been applied to simpler
versions of growing networks\cite{klr,doro,kr}.  When the attachment and
creation rates are given by Eq.~(\ref{rates}), the degree distribution
$N_{ij}(t)$ evolves according to the rate equations
\begin{eqnarray}
\label{Nij}
{d N_{ij}\over dt}&=&
(p+q)\left[{(i-1+\lambda)N_{i-1,j}-(i+\lambda)N_{ij}\over I+\lambda N}\right]\\
&+&q\left[{(j-1+\mu)N_{i,j-1}
-(j+\mu)N_{ij}\over J+\mu N}\right]
+p\,\delta_{i0}\delta_{j1}.\nonumber
\end{eqnarray}
The first group of terms on the right-hand side account for the changes in
the in-degree of target nodes.  These changes arise by simultaneous creation
of a new node and link (with probability $p$) or by creation of a new link
only (with probability $q$).  For example, the creation of a link to a node
with in-degree $i$ leads to a loss in the number of such nodes.  This occurs
with rate $(p+q)(i+\lambda)N_{ij}$, divided by the appropriate normalization
factor $\sum_{i,j}(i+\lambda)N_{ij}=I+\lambda N$.  The factor $p+q=1$ in
Eq.~(\ref{Nij}) has been written to make explicit the two types of relevant
processes.  Similarly, the terms in the second group of terms accounts for
changes in the out-degree.  These occur due to the creation of new links
between already existing nodes -- hence the prefactor $q$.  The last term
accounts for the continuous introduction of new nodes with no incoming links
and one outgoing link.  As a useful self-consistency check, we can easily
verify that the total number of nodes, $N=\sum_{i,j} N_{ij}$, obeys $\dot
N=p$, in agreement with Eq.~(\ref{NIJ}).  In the same spirit, the total in-
and out-degrees, $I=\sum_{i,j} iN_{ij}$ and $J=\sum_{i,j} jN_{ij}$, obey
$\dot I=\dot J=1$.

By solving the first few of Eqs.~(\ref{Nij}), it is clear that the $N_{ij}$
grow linearly with time.  Accordingly, we substitute $N_{ij}(t)=t\,n_{ij}$,
as well as $N=pt$ and $I=J=t$, into Eqs.~(\ref{Nij}) to yield a recursion
relation for $n_{ij}$.  Using the shorthand notations,
\begin{eqnarray*}
a=q\,{1+p\lambda\over 1+p\mu}\quad {\rm and}\quad 
b=1+(1+p)\lambda,
\end{eqnarray*}
the recursion relation for $n_{ij}$ simplifies to
\begin{eqnarray}
\label{nij}
[i+a(j+\mu)+b]n_{ij}
&=&(i-1+\lambda)n_{i-1,j}\nonumber\\
&+&a(j-1+\mu)n_{i,j-1}\nonumber\\
&+&p(1+p\lambda)\delta_{i0}\delta_{j1}.
\end{eqnarray}

We first consider the in-degree and out-degree distributions, ${\cal
  I}_i(t)=\sum_j N_{ij}(t)$ and ${\cal O}_j(t)=\sum_i N_{ij}(t)$.  Because of
the linear time dependence of the nodes degrees, we write ${\cal
  I}_i(t)=t\,I_i$ and ${\cal O}_j(t)=t\,O_j$.  The densities $I_i$ and $O_j$
satisfy
\begin{eqnarray}
\label{Ii}
(i+b)I_{i} &=&(i-1+\lambda)I_{i-1}+p(1+p\lambda)\delta_{i0},\\
\label{Oj}
\left(j+{1\over q}+{\mu\over q}\right)O_j 
&=&(j-1+\mu)O_{j-1}+p{1+p\mu\over q}\delta_{j1},
\end{eqnarray}
respectively.  The solution to these recursion formulae may be expressed
in terms of the following ratios of gamma functions
\begin{equation}
\label{I-sol}
I_{i}=I_0\,{\Gamma(i+\lambda)\,\Gamma(b+1)\over
\Gamma(i+b+1)\,\Gamma(\lambda)},
\end{equation}
with $I_0=p(1+p\lambda)/b$, and
\begin{eqnarray}
\label{O-sol}
O_{j}=O_1\,{\Gamma(j+\mu)\,\,\Gamma(2+q^{-1}+\mu q^{-1})\over 
\Gamma(j+1+q^{-1}+\mu q^{-1})\,\Gamma(1+\mu)},
\end{eqnarray}
with $O_1=p(1+p\mu)/(1+q+\mu)$.

{}From the asymptotics of the gamma function, the asymptotic behavior of
the in- and out-degree distributions have the power law forms, 
\begin{eqnarray}
\label{in}
I_i\sim i^{-\nu_{\rm in}},\qquad \nu_{\rm in}&=&2+p\lambda,\\
\label{out}
O_j\sim j^{-\nu_{\rm out}}, \qquad
\nu_{\rm out}&=&1+q^{-1}+\mu pq^{-1}.  
\end{eqnarray}
These exponents for the degree distributions constitute one of our primary
results.  Note that $\nu_{\rm in}$ depends on $\lambda$ (an in-degree
feature) while $\nu_{\rm out}$ depends on $\mu$ (an out-degree feature).
Notice also that both the exponents are greater than 2.

We can also solve the recursion relation (\ref{nij}) for $n_{ij}$ when $i$ or
$j$ is small.  For example, we can express $n_{i1}$ as the ratio of two gamma
functions.  Then we can express $n_{i2}$ as the sum of two such ratios, {\it
  etc}.  While there appears to be no simple general expression for the joint
distribution, we can extract the limiting behaviors of $n_{ij}$ when $i$ or
$j$ is large.  We find
\begin{equation}
\label{extreme}
n_{ij}\sim\cases{i^{-\xi_{\rm in}}\,j^{\mu}, & $1\ll j\ll i$;\cr
j^{-\xi_{\rm out}}\,i^{\lambda-1}, & $1\ll i\ll j$;}
\end{equation}
with
\begin{eqnarray}
\label{xis}
\xi_{\rm in} &=&\nu_{\rm in}+{q\over p}\,
{(\nu_{\rm in}-1)(\nu_{\rm out}-2)\over \nu_{\rm out}-1} \nonumber \\
\xi_{\rm out} &=&\nu_{\rm out}+{1\over p}\,
{(\nu_{\rm out}-1)(\nu_{\rm in}-2)\over \nu_{\rm in}-1}.
\end{eqnarray}
Thus the in- and out-degrees of a node are {\em correlated} -- otherwise, we
would have $n_{ij}=I_iO_j\sim i^{-\nu_{\rm in}}j^{-\nu_{\rm out}}$.  This
correlation between node degrees is our second basic result.  

The analytical form of the joint distribution greatly simplifies when
$\nu_{\rm in}=\nu_{\rm out}$, corresponding to $a=1$ and
$\mu+b=2\lambda$.  In this region of the parameter space, the recursion
relation (\ref{nij}) reduces to
\begin{eqnarray}
\label{nij*}
(i+j+2\lambda)n_{ij}&=&(i-1+\lambda)n_{i-1,j}\nonumber\\
                       &+&(j-1+\mu)n_{i,j-1} \nonumber\\
                       &+&p(1+p\lambda)\delta_{i0}\delta_{j1}.
\end{eqnarray}
Equation (\ref{nij*}) is simpler than the general recursion 
(\ref{nij}) since the node degrees $i$ and $j$ now appear with equal
prefactors.  This feature allows us to transform Eq.~(\ref{nij*}) into a
constant-coefficient recursion relation.  Indeed, the substitution
\begin{equation}
\label{mij}
n_{ij}={\Gamma(i+\lambda)\,\Gamma(j+\mu)\over 
\Gamma(i+j+2\lambda+1)}\,\,m_{ij}
\end{equation}
reduces (\ref{nij*}) to
\begin{equation}
\label{m}
m_{ij}=m_{i-1,j}+m_{i,j-1}+\gamma\,\delta_{i0}\delta_{j1},  
\end{equation}
with
$\gamma=p(1+p\lambda)\,{\Gamma(1+2\lambda)/(\Gamma(\lambda)\,\Gamma(\mu+1))}$.
We solve Eq.~(\ref{m}) by the generating function technique.  Multiplying
Eq.~(\ref{m}) by $x^iy^j$ and summing over all $i\geq 0, j\geq 1$ yields
\begin{equation}
\label{Mxy}
{\mathcal M}(x,y)\equiv\sum_{i=0}^\infty \sum_{j=1}^\infty m_{ij}x^iy^j 
={\gamma y\over 1-x-y}.
\end{equation}
Expanding this latter expression we obtain
\begin{equation}
\label{mij-sol}
m_{ij}=\gamma\,\,{\Gamma(i+j)\over \Gamma(i+1)\,\Gamma(j)}.
\end{equation}
Combining Eqs.~(\ref{mij}) and (\ref{mij-sol}) gives the joint in- and
out-degree distribution
\begin{equation}
\label{nij-sol}
n_{ij}=\gamma\,\,{\Gamma(i+\lambda)\,\Gamma(j+\mu)\,\Gamma(i+j)\over 
\Gamma(i+1)\,\Gamma(j)\,\Gamma(i+j+2\lambda+1)}.
\end{equation}
In analogy to Eq.~(\ref{extreme}), this joint distribution reduces to
\begin{eqnarray}
\label{nij-scal}
n_{ij}=\gamma\,{i^{\lambda-1}j^{\mu}\over (i+j)^{2\lambda+1}}.
\end{eqnarray}
in the limit $i\to\infty$ and $j\to\infty$.

Another manifestation of the correlation in the degree distribution becomes
evident by fixing the in-degree $i$ and allowing the out-degree $j$ to vary.
We find that $n_{ij}$ reaches a maximum value when $j=i\mu/[2+(1+p)\lambda]$
(here we consider large $i$ and assume that $\mu>0$).  Correspondingly, the
average out-degree always scales linearly with the in-degree, $\langle
j\rangle= i(\mu+1)/[(1+p)\lambda]$ (here the coefficient is always positive).
Thus popular nodes -- those with large in-degree -- also tend to have large
out-degrees.  A dual property also holds: Nodes with large out-degree --
those where many links originate -- also tend to be popular.

Let us now compare our predictions with empirical observations for the
world-wide web.  The relevant results for the node degrees are\cite{brod}
\begin{equation}
\label{empir}
\nu_{\rm in}\approx 2.1, \quad
\nu_{\rm out}\approx 2.7, \quad
{\cal D}_{\rm in}={\cal D}_{\rm out}\approx 7.5,
\end{equation}
Setting the observed value ${\cal D}_{\rm in}={\cal D}_{\rm out}=7.5$ to
$p^{-1}$ (see the discussion following Eq.~(\ref{NIJ})) we see that the
predictions (\ref{in})--(\ref{out}) match the observed values of the in- and
out-degree exponents when $\lambda=0.75$ and $\mu=3.55$, respectively.  With
these parameter values we also have $\xi_{\rm in}\approx 5.0$ and $\xi_{\rm
  out}\approx 3.9$.  Empirical measurements of these exponents would provide
a definitive test of our model.

We have also investigated a simplified model with node creation rate
$A_i=i+\lambda$, as above, but with link creation rate $C(j,i)=j+\mu$, which
does not depend on the popularity of the target node $i$ ({\em linear-linear}
rates).  For this model, the rate equations for the evolution of the number
of nodes with degrees $(i,j)$ have a similar structure to Eqs.~(\ref{Nij})
and they can be solved by the same approach as that given for the network
with linear-bilinear growth rates.  We find that the in- and out-degree
distributions again have power-law forms.  Moreover, the out-degree exponent
is still given by Eq.~(\ref{in}), while the value of the in-degree exponent
is now $\nu_{\rm in}=1+\lambda+p^{-1}$.  If we set $p^{-1}=7.5$ to reproduce
the correct average degree of the web graph, we see that $\nu_{\rm in}$ must
be larger than 8.5.  Similarly the linear-linear model with $A_i=i+\lambda$
and $C(j,i)=i+\mu$ gives a power-law in-degree distribution but the
exponential out-degree distribution $O_j=p^2q^{j-1}$.  Therefore
linear-linear rate models cannot match empirical observations from the web.

Parenthetically, we can also solve completely the growing network with both
constant node creation rate and constant link creation rate, $A_i=1$ and
$C(j,i)=1$.  While not necessarily a realistic model, it provides a useful
exactly solvable case.  By following the basic steps of the rate equation
approach, we find the joint distribution
\begin{equation}
n_{ij}= {p^2q^{j-1}\over{2^{i+j}}}\,\,{\Gamma(i+j)\over\Gamma(i+1)\Gamma(j)},
\end{equation}
from which we deduce the in- and out-degree distributions:
$I_i=p^2/(1+p)^{i+1}$ and $O_j=p^2q^{j-1}$. Again, the in- and out-degrees of
a node are correlated.

In summary, we have studied a growing network model which incorporates: (i)
node creation and immediate attachment to a pre-existing node, and (ii) link
creation between pre-existing nodes.  The combination of these two processes
naturally leads to non-trivial in-degree and out-degree distributions.  We
computed many structural properties of the resulting network by solving the
rate equations for the evolution of the number of nodes with given in- and
out-degree.  For link attachment rate linear in the target node degree and
also link creation rate linear in the degrees of the two end nodes, power-law
in- and out-degree distributions are dynamically generated.  By choosing the
parameters of the growth rates in a natural manner these exponents can be
brought into accord with recent measurements of the web.  Within this class
of models, the linear-bilinear growth rates appears to be a viable candidate
for describing the link structure of the web graph.  The model also predicts
power-law behavior when {\it e.g.}, the in-degree is fixed and the out-degree
varies.  Significant correlations between the in- and out-degrees of a node
develop spontaneously, in agreement with everyday experience.  Quantitative
measurements of correlations in the web graph would test our model and help
construct a more realistic model of the world-wide web.

\medskip\noindent We are grateful for financial support of this work from NSF
grant DMR9978902 and ARO grant DAAD19-99-1-0173 (PLK and SR), and a grant
from the EPSRC (GJR).

\end{multicols}
\end{document}